\documentclass[11pt,a4paper]{article}
\pdfoutput=1

\usepackage{jheppub}\usepackage{graphicx,placeins,amsmath,amssymb,subfigure,color,ulem}
\usepackage[applemac]{inputenc}
\usepackage{cancel}

\def\lsim{\raise0.3ex\hbox{$<$\kern-0.75em\raise-1.1ex\hbox{$\sim$}}}
\def\gsim{\raise0.3ex\hbox{$>$\kern-0.75em\raise-1.1ex\hbox{$\sim$}}}

\begin{document}

\title{Phase shifts in $I{=}2$ $\pi\pi$-scattering from two lattice approaches}

\title{Phase shifts in $I{=}2$ $\pi\pi$-scattering from two lattice approaches}

\author[a]{T.~Kurth}
\author[b]{N.~Ishii}
\author[c]{T.~Doi}
\author[d,e]{S.~Aoki}
\author[c,f]{T.~Hatsuda}

\affiliation[a]{Bergische Universit\"at Wuppertal, Gaussstr.\,20, D-42119 Wuppertal, Germany.}
\affiliation[b]{Kobe Branch, Center for Computational Sciences, University of Tsukuba, in RIKEN
Advanced Institute for Computational Science(AICS), PortIsland, Kobe 650-0047, Japan.}
\affiliation[c]{Theoretical Research Division, Nishina Center, RIKEN, Wako 351-0198, Japan.}
\affiliation[d]{Yukawa Institute for Theoretical Physics, Kyoto University, Kitashirakawa Oiwakecho, Sakyo-ku, Kyoto 606-8502, Japan}
\affiliation[e]{Center for Computational Sciences, University of Tsukuba, Tsukuba 305-8577, Japan.}
\affiliation[f]{Kavli IPMU, The University of Tokyo, Kashiwa 277-8583, Japan.}
\emailAdd{thorsten.kurth@uni-wuppertal.de}
\emailAdd{ishii@ribf.riken.jp}
\emailAdd{doi@ribf.riken.jp}
\emailAdd{saoki@yukawa.kyoto-u.ac.jp}
\emailAdd{thatsuda@riken.jp}

\abstract{
We present a lattice QCD study of the phase shift of 
$I{=}2$ $\pi\pi$ scattering on the basis of two different approaches:
the standard finite volume approach by Lüscher and the recently introduced HAL QCD potential method.
Quenched QCD simulations are performed on lattices with extents $N_s{=}16,24,32,48$ and $N_t{=}128$ as well as lattice spacing $a{\sim}0.115\,\mathrm{fm}$ and a pion mass of $m_\pi{\sim}940\,\mathrm{MeV}$.
The phase shift and the scattering length are calculated in these two methods.
In the potential method, the error is dominated by the systematic uncertainty
associated with the violation of rotational symmetry due to finite lattice spacing. In
Lüscher's approach, such systematic uncertainty is difficult to be evaluated and thus is not
included in this work. 
A systematic uncertainty attributed to the quenched approximation, however, is not evaluated in both methods.
In case of the potential method, the phase shift can be calculated
for arbitrary energies below the inelastic threshold.
The energy dependence of the phase shift is also obtained
from L\"uscher's method using different volumes and/or nonrest-frame extension of it.
The results are found to agree well with the potential method.
}


\maketitle

\section{Introduction}
Quantum chromodynamics, the theory of the strong interaction, is now known for about four decades and its predictive power is undisputed. 
 Today, in the context of present lattice QCD calculations, armed with modern algorithms 
 and peta-flops computing systems, it is becoming possible to put nuclear physics on the ground of QCD. 
One such approach 
 is based on effective theories in which hadrons are the fundamental degrees of freedom and the appearing low energy coefficients have to be matched to the underlying theories \cite{Bernard:1995hc,Epelbaum:2009ij}.
The standard approach within QCD involves the computation of multi-nucleon correlation functions and thus suffers from the combinatorial growth of the numbers of Wick-contractions which have to be computed 
\cite{Detmold:2010oq,Doi:2012xd,Detmold:2012tg,Gunther:2013xj}. 
Recently, the HAL QCD collaboration proposed an alternative method \cite{Ishii:2006ec,Aoki:2009ji}, 
where interaction kernels (potentials) are first calculated on the lattice and then low-energy observables such as nucleon-nucleon (NN)
scattering phase shifts are extracted by solving the Schr\"odinger equation employing these potentials.
This method, which we will call the HAL QCD method or the potential method in this paper, has been widely applied to various hadronic interactions~\cite{Nemura:2008sp,Inoue:2010es,Doi:2011gq,Inoue:2012fv} (also cf. the recent review \cite{Aoki:2012tk}).

The aim of this paper is to compare L\"uscher's approach \cite{Luscher:1986pf,Luscher:1990ux},
which has been widely used in literature ~\cite{Sharpe:1992pp,Fukugita:1994ve,Yamazaki:2004qb,Aoki:2005uf,Aoki:2002ny,Beane:2011sc,Dudek:2012gj},
and the potential method applied to the $I{=}2$ two-pion scattering problem, where more accurate data than in $NN$ system can be obtained. In principle, both methods should give the same results. In practice, however, the systematic uncertainties are different. Using our high precision data, we would like to address these uncertainties in detail in order to make pros and cons of both methods clear. 
We will study whether, where and how these methods are comparable, or either method is better than the other.

\section{L\"uscher's method}
L\"uscher's method allows to extract information on two-particle scattering amplitudes from the energy levels of the specific state in a finite volume. The energy levels, which are discrete on the finite volume lattice, are shifted with respect to those of the non-interacting theory. L\"uscher found \cite{Luscher:1986pf,Luscher:1990ux} that this deviation could be used to obtain information on the corresponding S-matrix element.

\begin{figure}[htb]
\centering
\subfigure[Trace-disconnected]{
\includegraphics[scale=0.4]{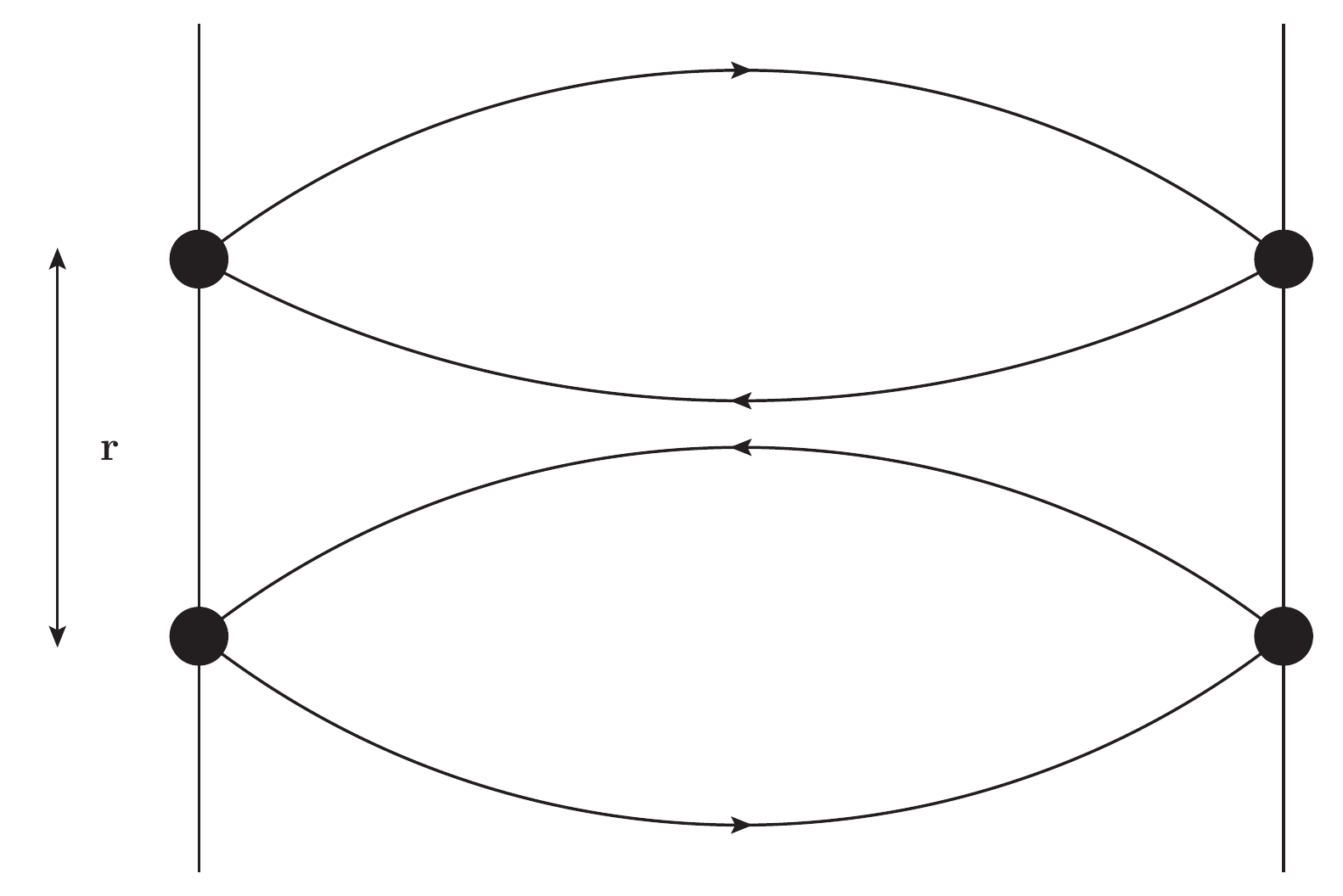}
}
\subfigure[Trace-connected]{
\includegraphics[scale=0.4]{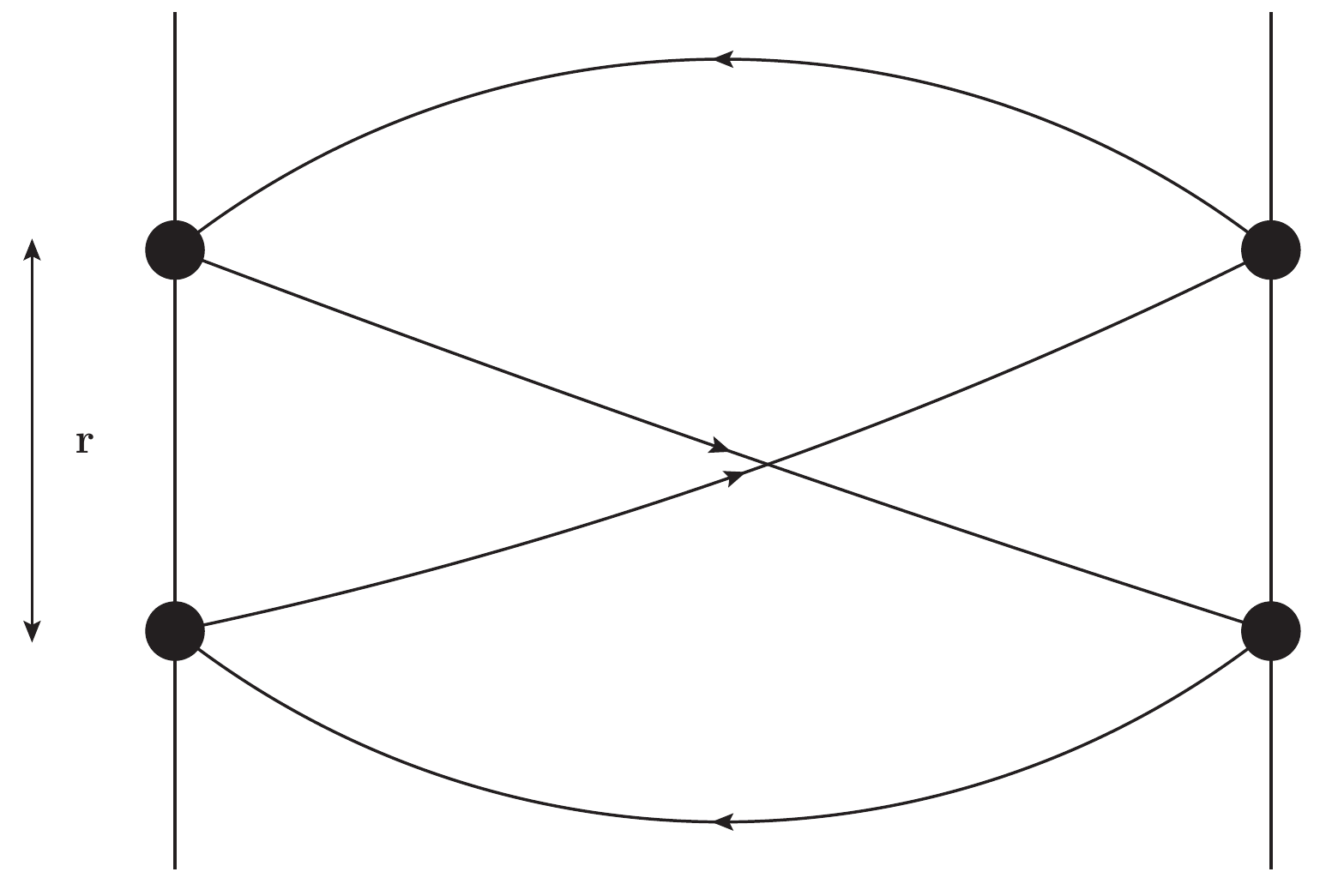}
}
\caption{\label{pipi_contractions}Contractions necessary to compute the $I{=}2$ two pion correlation function. The different contributions are referred to trace-disconnected (a) and trace connected (b) respectively.}
\end{figure}

For applying this method, we first have to compute the correlation function of the desired 2-particle state. In our case, this is the $\pi\pi$ scattering state in the $I{=}2$ channel and its correlation function is given by
\begin{equation}\label{corrfunc}
C_{\pi\pi}(t,t_0,\mathbf{r})\equiv\frac{1}{V}\sum\limits_\mathbf{x}\big\langle \pi^+(t,\mathbf{x}+\mathbf{r})\pi^+(t,\mathbf{x})J_{\pi^-}(t_0)J_{\pi^-}(t_0)\big\rangle,
\end{equation}
where $V=L^3$ is the spatial volume of the lattice.
We write $\pi^\pm$ in order to emphasize that we are considering the $I{=}2$ channel which does not contain disconnected diagrams. The contractions which have to be performed are displayed in Figure \ref{pipi_contractions}.

We consider the pseudo-scalar interpolating operator
$\pi^{+}=\bar{d}\gamma_5u$ and the source function $J_{\pi^-}$. 
In this work, we employ a wall source $J_{\pi^-}=W_{\pi^-}$, where
\begin{equation}\label{wall-source}
W_{\pi^-}(t_0)\equiv \sum\limits_{\mathbf{x},\mathbf{y}} \bar{u}(t_0,\mathbf{x})\gamma_5d(t_0,\mathbf{y}),
\end{equation}
as well as a Gaussian source $J_{\pi^-}=G_{r,\pi^-}$, where
\begin{equation}\label{gauss-source}
G_{r,\pi^-}(\mathbf{x}_0,t_0)\equiv\sum\limits_\mathbf{x, \mathbf{y}}e^{-\frac{(\mathbf{x}-\mathbf{x}_0)^2}{r^2}}e^{-\frac{(\mathbf{y}-\mathbf{x}_0)^2}{r^2}}\,
\bar{u}(t_0,\mathbf{x})\gamma_5d(t_0,\mathbf{y})
\end{equation}
with an appropriate radius $r$ (see below).
Furthermore, we need the single-pion correlation function given by
\begin{equation}
C_\pi(t,t_0)\equiv\sum\limits_{{\mathbf{x}}}\big\langle \pi^+(t,\mathbf{x})J_{\pi^-}(t_0)\big\rangle.
\end{equation}
Let us now consider the ratio
\begin{equation}\label{ratio-luscher}
\tilde{R}(t,t_0)\equiv \frac{\sum\nolimits_\mathbf{r} C_{\pi\pi}(t,t_0,\mathbf{r})}{C_{\pi}^2(t,t_0)}
\end{equation}
and the corresponding effective mass
\begin{equation}\label{energy_diff}
\Delta E(t)\equiv \frac{1}{2}\log\frac{\tilde{R}(t-1,t_0)}{\tilde{R}(t+1,t_0)}.
\end{equation}
For $t{\rightarrow}\infty$, (\ref{energy_diff}) approaches the (relativistic) energy difference between an interacting two-pion system and a non-interacting one. Thus,
\begin{equation}
\frac{\Delta E(t)}{2}\stackrel{t{\rightarrow}\infty}{\longrightarrow}\sqrt{k^2+m_\pi^2}-m_\pi,
\label{eq:Elimit}
\end{equation}
where $k$ is the asymptotic scattering momentum. The scattering phase shift $\delta(k)$ can be obtained by solving a nonlinear equation which involves the toroidal zeta function \cite{Luscher:1986pf}:
\begin{equation}\label{zetafunc}
k\cot\delta(k)=\frac{1}{\pi L}\sum\limits_{\mathbf{n}\in\mathbb{Z}^3}\frac{1}{|\mathbf{n}|^2-\tilde{k}^2},
\end{equation}
with $\tilde{k}\equiv kL/(2\pi)$ and box-size $L$. It is understood that the sum is regularized by analytic continuation and for numerical applications a convergence acceleration is used \cite{Luscher:1986pf,Luscher:1990ux}.
The scattering length $a_{\pi\pi}^{I{=}2}$ is defined by the low-energy limit
\begin{equation}
a_{\pi\pi}^{I{=}2}\equiv \lim\limits_{k\rightarrow 0}\frac{\tan\delta(k)}{k}.
\end{equation}
Therefore, one ideally has to compute $\delta(k)$ for many small values of $k$, which requires the use of large lattices (since $k{\sim}1/L$). Indeed we take a large lattice in this paper, so that
it is suitable to expand the energy level of the lowest lying continuum state in the $A_1$ representation of the cubic group \cite{Mandula:1983ut} for $L\gg a_{\pi\pi}^{I{=}2}$ via (cf. \cite{Luscher:1986pf,Beane:2003da})
\begin{equation}\label{deltaE0}
\Delta E_0=-\frac{4\pi a_{\pi\pi}^{I{=}2}}{m_\pi L^3}\left[1+c_1\frac{a_{\pi\pi}^{I{=}2}}{L}+c_2\left(\frac{a_{\pi\pi}^{I{=}2}}{L}\right)^2\right]+\mathcal{O}\left(\frac{1}{L^6}\right),
\end{equation}
with  $c_1{=}-2.837297$ and $c_2{=}6.375183$. 
Straightforward estimate with $L{\approx} 3.7\,\mathrm{fm}$ and $a_{\pi\pi}^{I{=}2}{\approx} 0.16\,\mathrm{fm}$ (see below)
suggests that the truncation error is negligible. We note, however, that there may exist a non-negligible error in eq.~(\ref{deltaE0}) attributed to the quenched approximation used in this study~\cite{Bernard:1995ez}.
In order to determine the scattering phase shift $\delta(k)$ at larger asymptotic momenta $k$, we use
 the moving-frame extension introduced by Rummukainen and Gottlieb \cite{Rummukainen:1995vs},
  which was already applied to the $I{=}2$ $\pi\pi$-system in \cite{Yamazaki:2004qb}.

\section{Potential method}
The potential method, which was originally introduced by the HAL QCD collaboration \cite{Ishii:2006ec,Aoki:2009ji}, also allows for extracting S-matrix elements from lattice correlation functions. The major difference to L\"uscher's finite volume method is that the interaction kernel (potential)  is extracted at short distances in the finite volume,  while the scattering phase shifts are calculated by solving the Schr\"odinger equation in the infinite volume. In practice, the potential method involves an additional expansion in terms of its non-locality with respect to the relative coordinate of the scattered particles. 
On the long run, the potential method will be used to solve multi-nucleon scattering problems and thus only heavy particles such as nucleons appear in the system. For that reason, we mainly focus on heavy pions  in this work.

In \cite{Aoki:2009ji} it was shown that the Nambu-Bethe-Salpeter (NBS) wave function satisfies the time-independent Schr\"odinger equation:
\begin{equation}\label{se-equation}
\left(\frac{\mathbf{k}^2}{m_\pi}-H_0\right)\psi_\mathbf{k}(\mathbf{r})=\int\mathrm{d}^3r'\,U(\mathbf{r},\mathbf{r}')\psi_\mathbf{k}(\mathbf{r}'),
\end{equation}
where $H_0\equiv - \nabla_{\mathbf{r}}^2/m_\pi$ is the non-interacting part of the two-pion Hamiltonian. Note that the non-local potential $U(\mathbf{r},\mathbf{r}')$ is independent of $\mathbf{k}$ \cite{Krolikowski:1956,Aoki:2009ji}. Here the equal-time NBS wave function can be written as
\begin{equation}\label{NBS-WF}
\psi_\mathbf{k}(\mathbf{r})\equiv \frac{1}{V}\sum\limits_{\mathbf{x}}\langle 0\vert\pi^+(\mathbf{x}+\mathbf{r})\pi^+(\mathbf{x})\vert \pi^+(\mathbf{k})\pi^+(-\mathbf{k})\rangle_{\rm in},
\end{equation}
where $\vert \pi^+(\mathbf{k})\pi^+(-\mathbf{k})\rangle_{\rm in}$ is a two-pion instate with relative momentum $\mathbf{k}$.
We can now relate the two-pion correlation function (\ref{corrfunc}) to the NBS wave function (\ref{NBS-WF}) via
\begin{equation}\label{WF-CF}
C_{\pi\pi}(\mathbf{r},t,t_0)=\sum\limits_\mathbf{k} \psi_\mathbf{k}(\mathbf{r})\,a_{\mathbf{k}}\,e^{-E_\mathbf{k} (t-t_0)},
\end{equation}
with Fourier coefficients $a_\mathbf{k}={}_{\rm in}\langle\pi^+(\mathbf{k})\pi^+(-\mathbf{k})\vert J_{\pi^-}(0)J_{\pi^-}(0)|0\rangle$ and $E^2_\mathbf{k}=4(\mathbf{k}^2+m_\pi^2)$.
We perform a derivative expansion of the non-local potential 
 and only keep the leading-order term. For the two-pion system, we can thus write
\begin{equation}\label{veloexpand}
U(\mathbf{r},\mathbf{r}')=V_C(\mathbf{r})\,\delta(\mathbf{r}-\mathbf{r}')
+\mathcal{O}(\nabla_{\mathbf{r}}^2),
\end{equation}
where $V_C$ is central potential and only dependent on $r{=}|\mathbf{r}|$.
Combining (\ref{veloexpand}) with (\ref{se-equation}) and (\ref{WF-CF}) we obtain \cite{Aoki:2009ji}
\begin{equation}\label{stationary-pot}
V_C(r)=\frac{\mathbf{k}^2}{m_\pi}+\frac{1}{m_\pi} \lim\limits_{(t-t_0)\rightarrow\infty} \frac{\nabla_{\mathbf{r}}^2 C_{\pi\pi}(\mathbf{r},t,t_0)}{C_{\pi\pi}(\mathbf{r},t,t_0)},
\end{equation}
where the Laplacian $\nabla_{\mathbf{r}}^2$ acts on the spatial components of $C_{\pi\pi}(\mathbf{r},t,t_0)$.

Both eq.~(\ref{eq:Elimit}) in L\"uscher's method and eq.~(\ref{stationary-pot}) in the potential method 
require data for large $t$ to guarantee the ground state saturation of the correlation function.
However, the signal-to-noise ratio becomes worse at large $t$, particularly for many-nucleon systems.
To overcome this difficulty, a generalized potential method has been introduced in \cite{HALQCD:2012aa} 
 on the basis of the  time-dependent Schr\"{o}dinger-like equation:
 We multiply (\ref{se-equation}) with $a_\mathbf{k}e^{-E_\mathbf{k}(t-t_0)}/C_\pi(t,t_0)^2\equiv\tilde{a}_\mathbf{k}e^{-\Delta E_\mathbf{k}(t-t_0)}$ and sum both sides over $\mathbf{k}$, obtaining
\begin{equation}\label{schroed}
\sum\limits_\mathbf{k} \tilde{a}_\mathbf{k}e^{-\Delta E_\mathbf{k}(t-t_0)}\left(\frac{\mathbf{k}^2}{m_\pi}-H_0\right)\psi_\mathbf{k}(\mathbf{r})=\int\mathrm{d}^3r'\,U(\mathbf{r},\mathbf{r}')R(\mathbf{r}',t),
\end{equation}
where we defined, analogous to (\ref{ratio-luscher}),
\begin{equation}\label{ratio-potential}
R(\mathbf{r},t,t_0)\equiv\frac{C_{\pi\pi}(\mathbf{r},t,t_0)}{C_\pi^2(t,t_0)}
\end{equation}
and $\Delta E_\mathbf{k}\equiv E_\mathbf{k}-2m_\pi$. It is straightforward to check that
\begin{equation}
\frac{\mathbf{k}^2}{m_\pi}e^{-\Delta E_\mathbf{k}(t-t_0)}=\left(-\frac{\partial}{\partial t}+\frac{1}{4m_\pi}\frac{\partial^2}{\partial t^2}\right) e^{-\Delta E_\mathbf{k}(t-t_0)}.
\end{equation}
Hence we find
\begin{equation}
\left(-\frac{\partial}{\partial t}+\frac{1}{4m_\pi}\frac{\partial^2}{\partial t^2}-H_0\right)R(\mathbf{r},t,t_0)=\int\mathrm{d}^3r'\,U(\mathbf{r},\mathbf{r}')R(\mathbf{r}',t,t_0).
\end{equation}
Since all the two-pion scattering states with different $\mathbf{k}$ 
obey the same non-local potential,  
 the ground state saturation with asymptotically large $t$
  is  no longer necessary in this time-dependent method.
The only condition for $t$ is that it should be large enough so that both inelastic contributions in the numerator of $R$  and the contribution from excited pions in 
the denominator of $R$ are negligible. The second condition is 
essentially the same condition for $t$ to extract the pion mass from $C_{\pi}(t,t_0)$.
The milder condition for $t$ is particularly  a good news for the two-nucleon system. 
At lowest order of the derivative expansion,
 this results in a modified expression for the local potential (cf. \cite{HALQCD:2012aa}):
\begin{equation}\label{time-dep-pot}
V_C(r)=\frac{1}{m_\pi}\frac{\nabla_\mathbf{r}^2 R(\mathbf{r},t,t_0)}{R(\mathbf{r},t,t_0)}
-\frac{\frac{\partial}{\partial t}R(\mathbf{r},t,t_0)}{R(\mathbf{r},t,t_0)}+\frac{1}{4m_\pi}\frac{\frac{\partial^2}{\partial t^2}R(\mathbf{r},t,t_0)}{R(\mathbf{r},t,t_0)}.
\end{equation}
In this improved method, one can also check the validity of the derivative expansion
 and improve the potential through the $t$-dependence of the right hand side of (\ref{time-dep-pot}).

Once the local potential $V_C(r)$ is obtained, it can be fitted by an empirical formula or, alternatively, interpolated. Since we need the potential only as input for the continuum Schr\"odinger equation, we perform a barycentric rational interpolation \cite{Floater:2007:BRI} to obtain a continuous function on $r$. Furthermore, we set the potential to zero for distances $r{>}r_\mathrm{cut}$, where $r_\mathrm{cut}$ is the largest distance available in the considered data set. This cut is far less aggressive than it might appear, since the measured potentials are essentially compatible with zero for distances $r{>}2\,\mathrm{fm}$. 

The interpolated potentials are then plugged into the radial part of the Schr\"odinger equation (\ref{se-equation}) with eq.~(\ref{veloexpand}) for zero orbital angular momentum $L{=}0$,
 which can be solved with respect to $\psi_k(r)$ for arbitrary $k^2$. The scattering phase $\delta(k)$ at a given asymptotic momentum $k$, can be obtained by noting that the asymptotic form of $\psi_k$ is given by a free radial wave \cite{Lin:2001ek,Aoki:2005uf,Aoki:2009ji}:
\begin{equation}
\psi_k(r)\stackrel{r\rightarrow \infty}{\longrightarrow} e^{i\delta(k)}\big(\cos\delta(k)\,j_0(kr)-\sin\delta(k)\,n_0(kr)\big),
\end{equation}
with the zeroth-order spherical Bessel and Neumann functions $j_0(r){=}\sin r/r$ and\\ $n_0(r){=}-\cos r/r$ respectively.
Therefore, the scattering phase shift can be obtained by taking the long distance limit of the following ratio:
\begin{equation}\label{delta-potential}
\tan\delta(k)=\left.\frac{kr\,j_0'(kr)-f_k(r)\,n_0'(kr)}{kr\,j_0(kr)-f_k(r)\,n_0(kr)}\right|_{r\rightarrow \infty},
\end{equation}
where
\begin{equation}
f_k(r)\equiv r\frac{\mathrm{d}\ln\psi_k(r)}{\mathrm{d}r}.
\end{equation}
Due to confinement the two-pion potential is localized and thus the asymptotic region is reached quickly. This also implies, that finite volume corrections in this method are expected to be very small \cite{Aoki:2009ji}.

In the final step, the scattering length and other phenomenologically relevant parameters such as the effective range can be computed by fitting the scattering phases to the effective range expansion \cite{Luscher:1990ux}:
\begin{equation}\label{ere}
\frac{k\,\cot\delta(k)}{m_\pi}=\frac{1}{m_\pi a_{\pi\pi}^{I{=}2}}+\frac{1}{2}m_\pi r_{\rm e}
\left(\frac{k^2}{m^2_\pi}\right)+P\,(m_\pi r_e)^3\left(\frac{k^2}{m^2_\pi}\right)^2+\mathcal{O}\left(\left(\frac{k^2}{m^2_\pi}\right)^3\right),
\end{equation}
where $r_{\rm e}$ 
is called the effective-range and $P$ the shape parameter. Both quantities parameterize the $k^2$ dependence of the phase shift $\delta(k)$ at low energies. 

\section{Lattice setup}\label{lattice-setup}
We generate approximately $400$ statistically independent quenched configurations using the Wilson plaquette action at $\beta{=}5.8726$ corresponding to $a{\sim}0.115\,\mathrm{fm}$. For the spacial size of the lattice, we employ $L_s^3{=}32^3$, for which we perform extensive analyses for the systematic uncertainties. We also study $L_s^3{=}16^3$, $24^3$ and $48^3$, for which the main part of the analysis, i.e., the analysis with the spatial wall source and temporal Dirichlet boundary condition, are performed.
The temporal extent is always chosen to be $N_t{=}128$. This amounts to spatial extents between $1.8\,\mathrm{fm}{\leq} L_s{\leq} 5.5\,\mathrm{fm}$. In the valence sector, we use 2 HEX smeared, tree-level clover improved Wilson quarks \cite{Capitani:2006ni,Durr:2010vn,Durr:2010aw,Durr:2011ap,Borsanyi:2012bs}, while the unimproved plaquette action is used for gauge fields. Smeared clover-Wilson actions have small cutoff effects on spectral quantities as shown in \cite{BMWScaling,Durr:2010aw}.
Since our ultimate goal is to compute multi-nucleon scattering, we tune $m_\pi$ to ${\sim}940\,\mathrm{MeV}$.
An exploratory study at smaller pion masses is also given in Appendix.~\ref{sec:mass-dep}.
Furthermore, we employ different combinations of source-types and boundary conditions in time direction. Whereas the spatial boundary conditions are always periodic, we use either anti-periodic or Dirichlet boundary conditions in time-direction. We further use wall (\ref{wall-source}) and Gaussian (\ref{gauss-source}) sources, where in the latter case the spatial location of one pion is chosen randomly and the second pion was separated by $L/2a\times (1,1,1)$ in lattice units. This symmetric setup claims to improve the overlap with the $J{=}0$ ground state \cite{Csikor:2006jy}. The size of the sources is set to $r{\sim}0.32\,\mathrm{fm}$, as it was done in previous studies \cite{Durr:2008zz}. To improve statistics, we employ four sources per config separated by 32 time-slices, where the Dirichlet boundary is always separated from the source by $T/2$.

We remark possible issues caused by the quenched approximation in this study.
As is well known, a quenched theory is not unitary.
Since both  L\"uscher's method and the potential method rely on the unitarity of the S-matrix,
the``phase shifts'' extracted from these methods
could loose their equivalence
due to quenching artifacts.
We expect, however, that our choice of the very heavy quark mass in this study 
would suppress the pathology in a quenched theory, so that
the equivalence of the phase shift from two methods is effectively recovered.
This is actually observed in our study, as will be shown in later sections.
Of course, more rigorous treatment requires full QCD simulations,
which is beyond the scope of this study.

\section{Error treatment}\label{error-treatment}
The statistical error is always computed by repeating the analysis on 2000 bootstrap samples. In order to estimate the overall systematic uncertainty, we use the histogram method \cite{Durr:2008zz,Durr:2010vn,Durr:2010aw,Durr:2011mp,Durr:2011ap,Borsanyi:2012bs}. We identified the following sources of systematic uncertainties: 

\begin{itemize}
\item \emph{Excited states:}
Possible contributions are estimated by varying the fitting range of the correlation functions. We use three fit ranges within $t_\mathrm{min}/a=15$ and $t_\mathrm{max}/a=48$:
\begin{itemize}
\item $R_1=\{t/a\in[15,48]\}$
\item $R_2=\{t/a\in[24,48]\}$
\item $R_3=\{t/a\in[33,48]\}$.
\end{itemize}
Below $t_\mathrm{min}$ we could not definitely exclude excited state contributions in L\"uscher's method. We consistently use the same ranges for fitting the effective energy in L\"uscher's method (\ref{energy_diff}) and for extracting the potential (\ref{time-dep-pot}).  The latter was averaged over the corresponding time-slices. This tests both methods with respect to their dependence on the ground-state saturation. 

\item\emph{Violation of rotational invariance:}
Rotational invariance is violated at short distance by the finite lattice cut-off
and at long distance by the finite lattice volume.
For the potential method, we estimate the resulting systematic uncertainty by extracting the potential for distances ${r}$ lying on-axis, 
on a plane-diagonal or on the cubic-diagonal and performing the full analysis on each of these choices separately.
We found that this uncertainty, particularly at short distance, is  the dominant source of the systematic error.
For L\"uscher's method, the systematic uncertainty induced by the violation of rotational invariance is difficult to estimate reliably,
since the spatial information is lost. We therefore do not include this specific systematic error in our analysis for Lüscher's method.
However, in both approaches, effects of violation of rotational invariance will be suppressed for smaller lattice spacings and for larger lattice volumes.

\item\emph{Interpolation and Fit:}
Before the Schr\"odinger equation is solved, the potential has to be interpolated or fitted by an empirical ansatz; we use a barycentric rational interpolation which is able to sample both the repulsive core and the long-distance tail. For testing the quality of this interpolation, we also performed empirical fits to Yukawa-type models or applied Gaussian process regressions. We have observed a good agreement in $\delta(k)$ among all three approaches. Thus for simplicity, we only kept the barycentric rational interpolation in our final analysis. The potential is compatible with zero at largest observed distances $r{=}R_\mathrm{max}$. Therefore, we set $V(r){=}0$ for $r{>}R_\mathrm{max}$. Assuming an exponentially fast decaying tail instead yields compatible results.

\item\emph{Asymptotic regime:}
By using (\ref{delta-potential}) for calculating the scattering phase shift, an additional systematic error can appear: it is attributed to the uncertainty of whether the asymptotic region in $r$ is reached or not. Therefore, we average (\ref{delta-potential}) over five disjoint regions of $r$ and propagate the differences among these regions into our systematic error.

\item\emph{Effective range expansion:}
When the scattering length and effective range is computed from the effective range expansion, possible higher order corrections in $k^2/m_{\pi}^2$ also contribute to the systematic error. This contribution is estimated by applying the effective range expansion fit two times: the first one is made with a conservative fit window in $E_\mathrm{CM}$ and only the first two terms of (\ref{ere}). In the second fit, the window is extended and the third term of (\ref{ere}) is also included.
\end{itemize}
Altogether we perform $3{\cdot}3{\cdot}5{\cdot} 2{=}90$ analyses for the potential method and $3{\cdot}2{=}6$ different analyses for L\"uscher's method in our final analysis. 
Additionally, we use $2$ different boundary conditions and $2$ different source types. The systematic uncertainties attributed to these effects are small and rather compared directly than propagated into our final systematic error using the histogram method.

The results obtained from the different analyses are collected in a histogram and weighted by their quality of fit $Q$. For computing the overall systematic uncertainty of the scattering phase $\delta(k)$ we use a unit weight whereas in case of the scattering length we apply the fit-quality obtained by the effective range expansion fit (we have checked that choosing a unit weight instead yields compatible results, i.e. the central value changes by less than $0.1\sigma$ and the systematic error increases by less than $1\%$). In all cases, the median gives our central value and the central $68\%$ the systematic error.


\section{Results}

\begin{figure}
\centering
\includegraphics[scale=1.1]{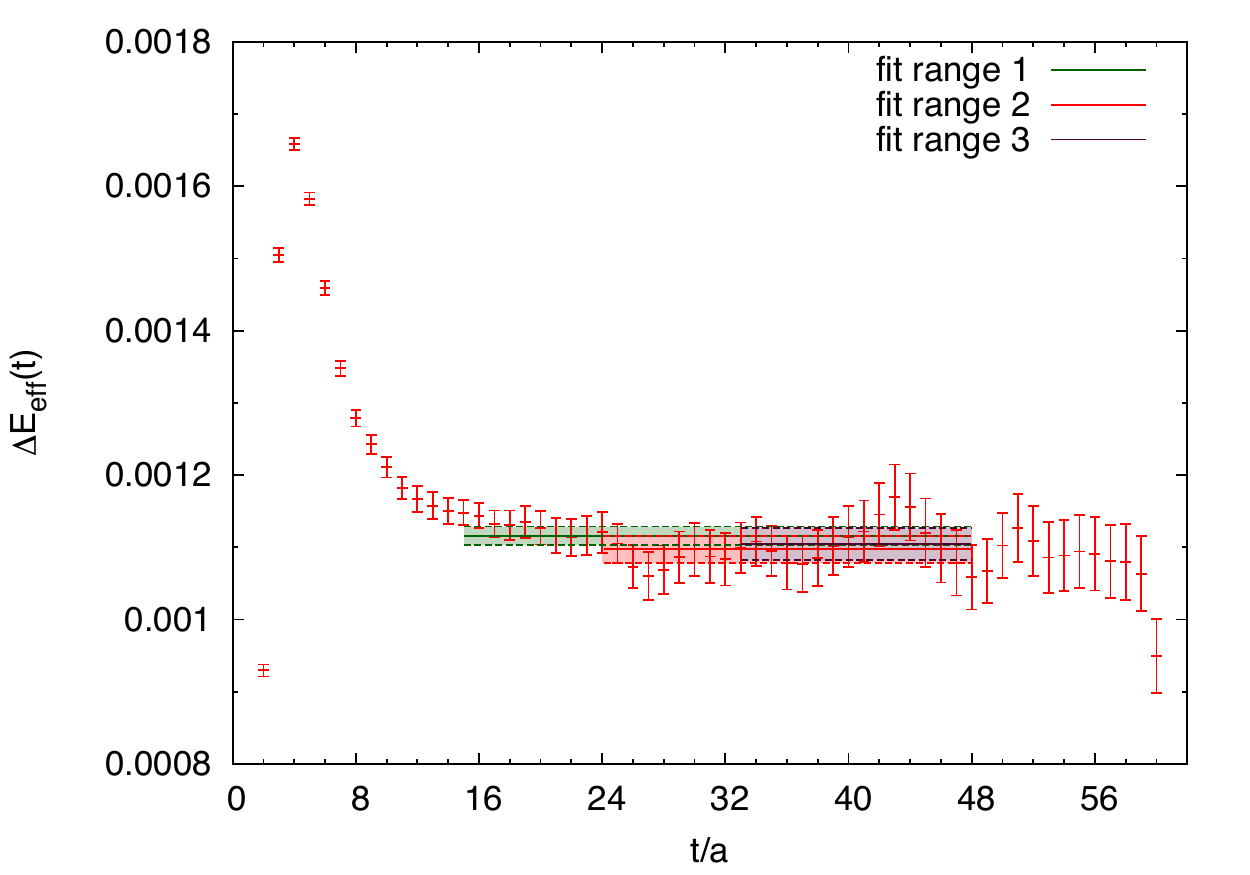}
\caption{\label{luscher-wall-dirichlet-plateau} Constant fits to effective mass plateau (\ref{energy_diff}) for the wall/Dirichlet data. Bands indicate the corresponding errors from the fits. Note that the systematic uncertainty attributed to the choice of the fitting range ($R_1,R_2,R_3$) is small ($L_s{=}3.7\,\mathrm{fm}$).}
\end{figure}

\begin{figure}
\centering
\includegraphics[scale=1.1]{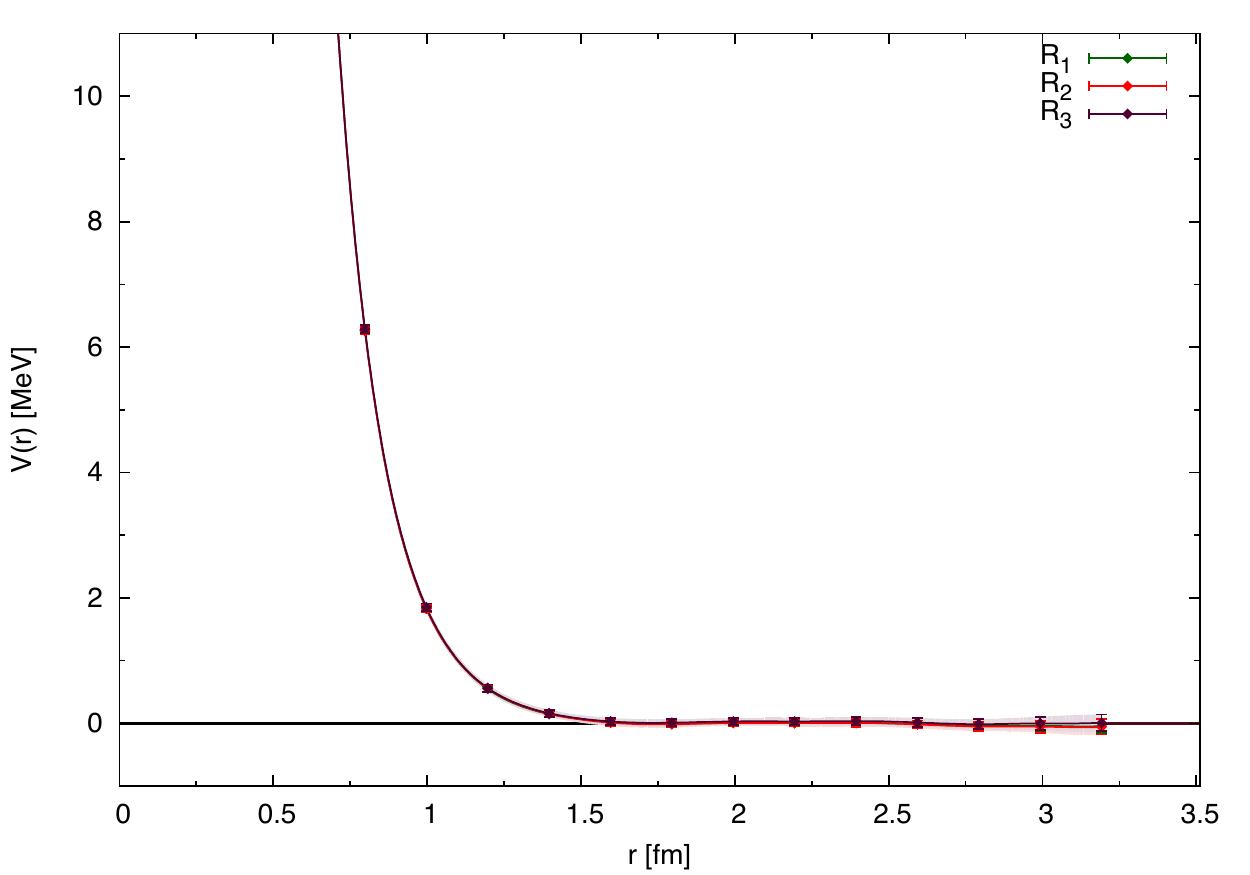}
\caption{\label{potential_diag3_wall} Two-pion potential for the different ranges $R_i$ computed on the cubic diagonal (points). The curves represent the results from the interpolation ($L_s{=}3.7\,\mathrm{fm}$).}
\end{figure}

\begin{figure}
\centering
\includegraphics[scale=1.1]{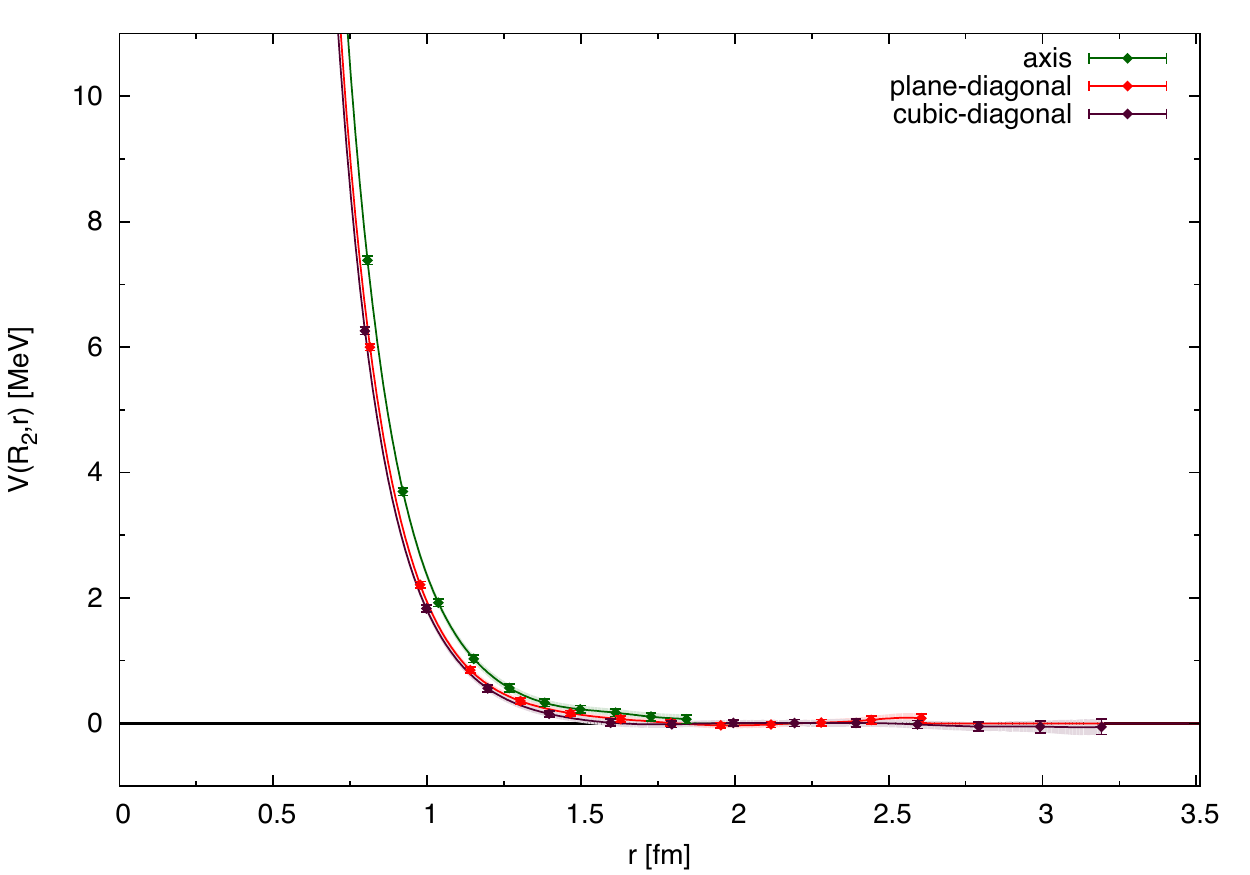}
\caption{\label{potential_range2_wall} Two-pion potential for $R_2$ and evaluated on axis (green), plane-diagonals (red) and cubic-diagonal (purple) and their corresponding interpolations ($L_s{=}3.7\,\mathrm{fm}$).}
\end{figure}

\begin{figure}
\centering
\includegraphics[scale=1.1]{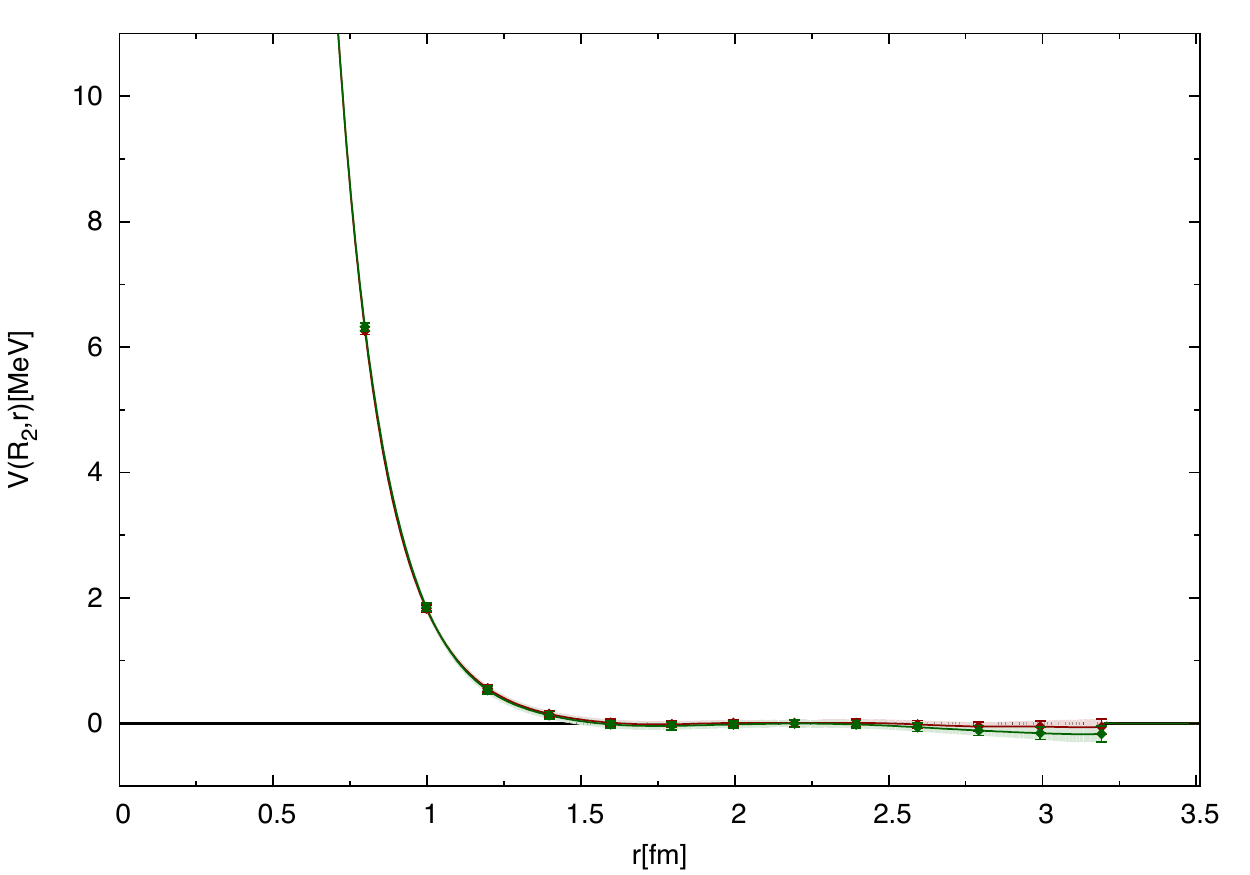}
\caption{\label{gauss-wall-dirichlet-compare} Source dependence of potential for $R_2$. The potential obtained from Gaussian sources (circles) is consistent with that obtained by employing wall sources (squares) ($L_s{=}3.7\,\mathrm{fm}$).}
\end{figure}

\begin{figure}
\includegraphics[scale=1.05]{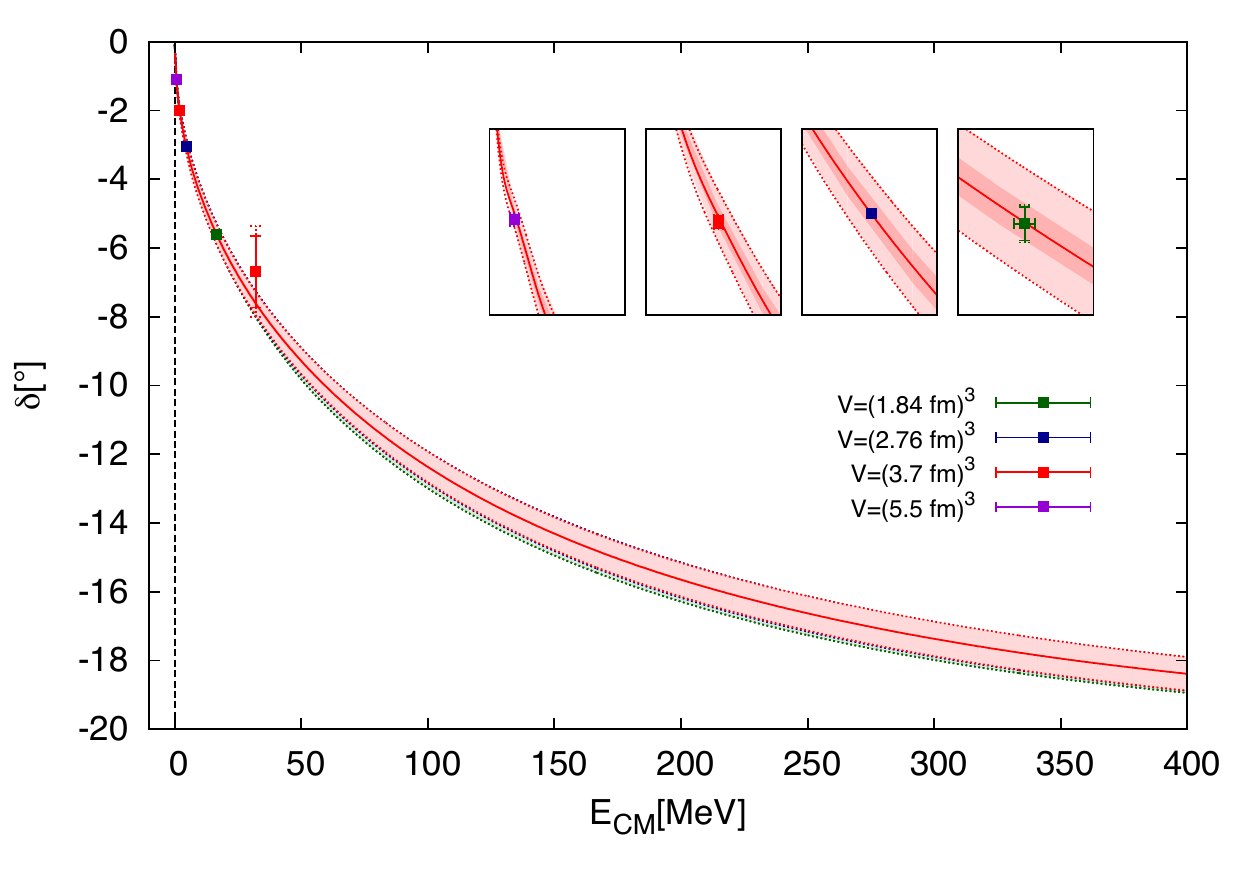}
\includegraphics[scale=1.05]{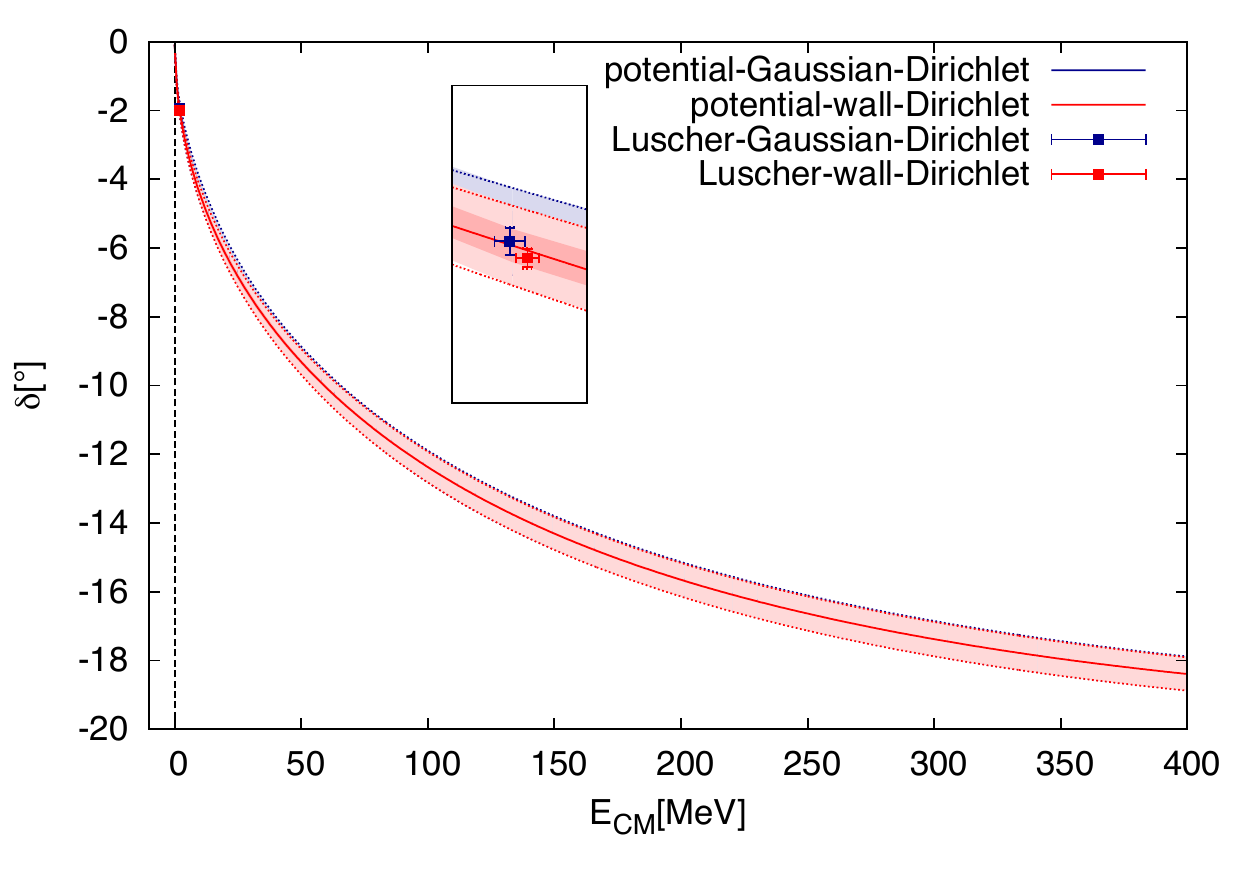}
\caption{\label{phases_dirichlet}Scattering phase shifts obtained from the potential and L\"uscher's method for wall (top) and wall and Gaussian sources (bottom). In the upper panel, the red band is obtained by the HAL QCD method 
using the potential obtained from $L_s{=}3.7\,\mathrm{fm}$. The green band is obtained from $L_s{=}1.8\,\mathrm{fm}$, and almost overlaps with one from $L_s{=}3.7\,\mathrm{fm}$. The point data are obtained by L\"uscher's method
with the center-of-mass frame on each volumes, except for the red point around $E_\mathrm{cm} \sim 30\,\mathrm{MeV}$, which is obtained on the $L_s{=}3.7\,\mathrm{fm}$ volume by applying L\"uscher's method to boosted system
with center-of-mass momentum $P_\mathrm{cm}{=}2\pi/L_s$. In the lower panel, we only used one volume and no boosted frames for the Gaussian source, so that only one data point can be displayed.}
\end{figure}

\begin{figure}
\centering
\includegraphics[scale=1.0]{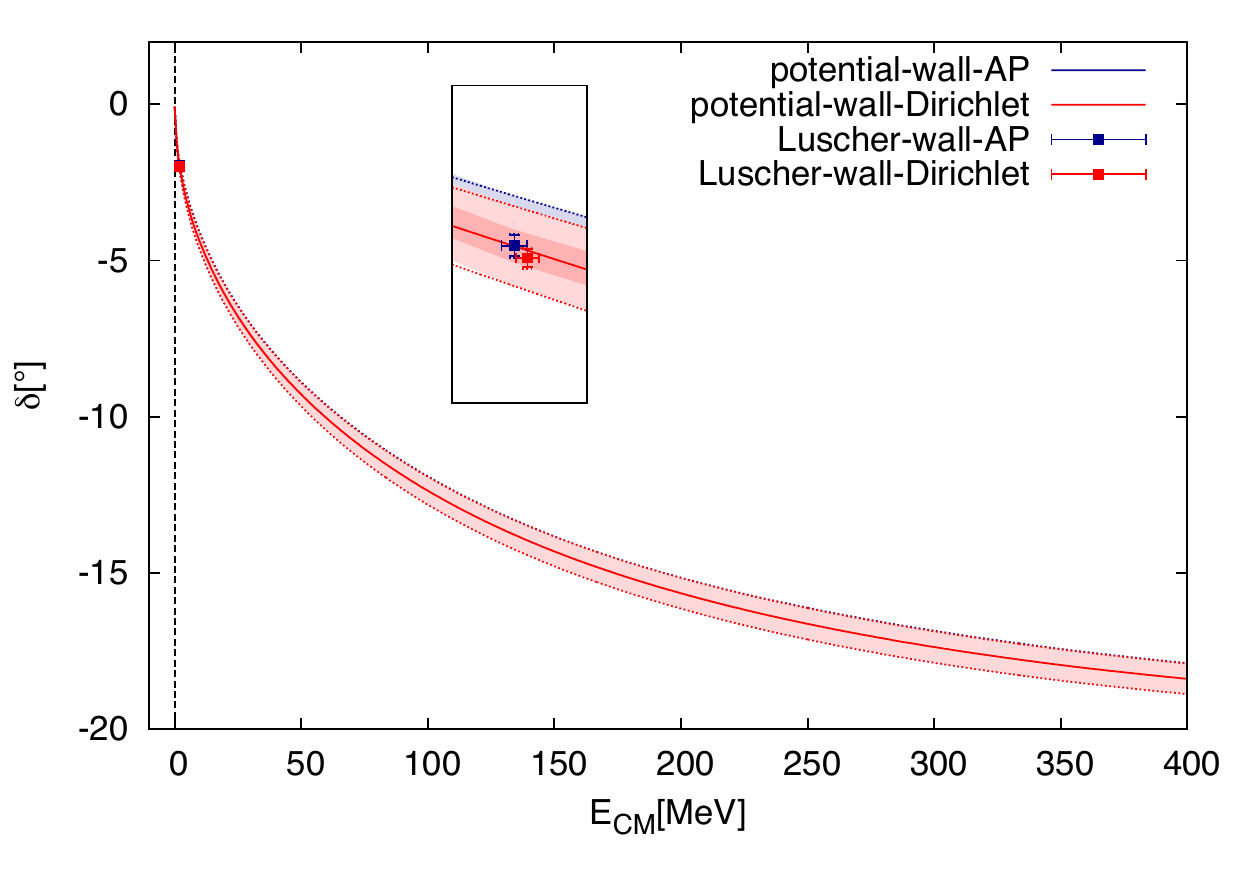}
\caption{\label{phase_compare_wall_dirichlet_antiper}Dependence of the scattering phase shift on the temporal boundary conditions. The plot shows $\delta$ vs. $E_\mathrm{CM}$ for the wall-source-data with anti-periodic or Dirichlet boundary conditions ($L_s{=}3.7\,\mathrm{fm}$).}
\end{figure}

\begin{figure}
\centering
\includegraphics[scale=1.2]{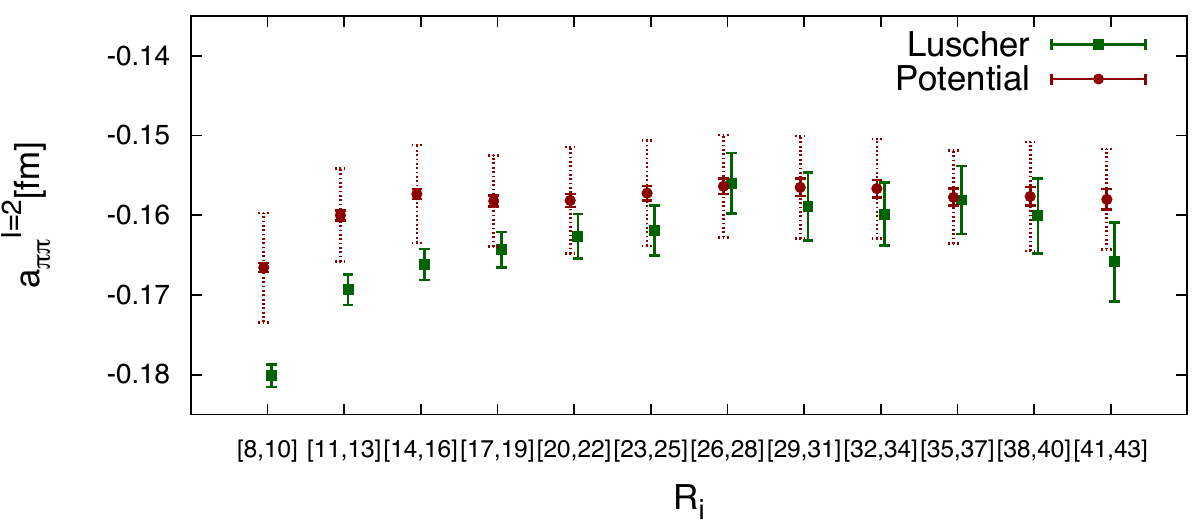}
\caption{\label{fig:t-dependence}Scattering lengths obtained by the potential or L\"uscher's method in dependence on the temporal range used for extracting the potential or $k^2$ ($L_s{=}3.7\,\mathrm{fm}$).}
\end{figure}

We are going to discuss the results of our analysis in detail now, focusing on a few possible sources of systematic uncertainties at a time.

\subsection{Source- and boundary-dependence}
For the first test, we compare the Gaussian and wall sources with Dirichlet boundary conditions at $m_\pi{\sim}940\,\mathrm{MeV}$ for both methods. The wall source is commonly used in conjunction with this type of boundary conditions \cite{Aoki:2009ji,Inoue:2010es,Doi:2011gq,Inoue:2012fv,HALQCD:2012aa}. The pion masses are extracted from the single-pion correlation functions by applying exponential fits. The energy shift $\Delta E$  is obtained by fitting the corresponding effective mass plateau (\ref{energy_diff}) to a constant as shown for the wall source in Figure \ref{luscher-wall-dirichlet-plateau}. From that value and $m_\pi$, the asymptotic momentum $k$ can be computed and plugged into (\ref{zetafunc}), yielding $\delta(k)$.
We obtain the lattice potential by computing  (\ref{time-dep-pot}), where the spatial and temporal derivatives in the right hand side have been evaluated numerically by symmetric five-point formulas.
Figure \ref{potential_diag3_wall} depicts the $r$ dependence of the potential obtained from the wall/Dirichlet data lying on the cubic diagonals. The plot includes all time slices ranging from $t_\mathrm{min}$ to $t_\mathrm{max}$ along with the interpolations. A good agreement among potentials from (\ref{time-dep-pot}) suggests that contributions from higher order terms in the derivative expansion are small.
The situation is different for the effect of rotational invariance violation: Figure \ref{potential_range2_wall} shows the potential on $R_2$ for on-axis, plane-diagonal and cubic-diagonal data points. Here we observe stronger deviations from the on-axis data compared to the other two. This uncertainty induced by rotational invariance violation is the major contribution to the systematic uncertainties of the potential method.
Figure \ref{gauss-wall-dirichlet-compare} depicts the source dependence of the potential by comparing the potential obtained from the two sources. The largest differences can be found in the tail, but everything is well compatible within errors. Therefore, observables sensitive to the asymptotic behavior of the potential - such as the scattering length - are only mildly affected by the choice of the source.
This suggests that contributions form higher order terms in the derivative expansion are small.
The fitted potential can be used to solve the Schr\"odinger equation (\ref{se-equation}) to obtain the two-pion wave functions in infinite volume. When combined with (\ref{delta-potential}), we are able to compute the scattering phase shift $\delta(k)$ for arbitrary $k$. The corresponding curves can be compared to the value computed by using L\"uscher's method. This is done in Figure \ref{phases_dirichlet} (upper panel) where the bands correspond to the results obtained from the potential method and the points correspond to those obtained from L\"uscher's method.
We observe that the results of both methods agree very well. In the potential method, the red band corresponds to the results obtained from $L_s{=}3.7\,\mathrm{fm}$. 
The same curve is drawn for $L_s{=}1.84\,\mathrm{fm}$ in green and highly agrees with the results obtained at $L_s{=}3.7\,\mathrm{fm}$, since only a tiny fraction of it can be seen at the lower edge of the red band. Comparing the potential on different volumes, we find that the finite volume artifact in the potential is negligible; all potentials agree within errors.
As a result, the phase shifts obtained from potentials computed on $L_s{=}(1.8-5.5)\,\mathrm{fm}$ agree very well within errors.
In L\"uscher's method, the phase shifts at different energies are obtained in the center-of-mass frame by changing the lattice spatial volume.
We also study the non-rest frame extension of L\"uscher's method for $L_s{=}3.7\,\mathrm{fm}$, which corresponds to the data point at $E_\mathrm{CM}{\sim} 30\,\mathrm{MeV}$.
The error bar is rather large since it is extracted in a boosted system. The lower panel of Figure \ref{phases_dirichlet} compares the phase shifts obtained using Gaussian and wall sources on our reference volume with spatial extent $L_s{=}3.7\,\mathrm{fm}$. A good agreement indicates that the source dependence is very mild, thus concluding that contributions form higher order terms in the derivative expansion are insignificant. 
Errors in the potential method are mainly due to violation of rotational invariance, whereas this systematic uncertainty has not been accounted for in Lüscher's method (cf. section \ref{error-treatment}).
We observe that the potential method not only gives a value consistent with the L\"uscher's method at particular $E_{\rm CM}$ but also provides a powerful tool for mapping out the dependence of the scattering phase $\delta(k)$ on the wide range of $E_\mathrm{CM}$ (Note that the data points on that curve are highly correlated, since they are all obtained deterministically by solving the Schroedinger equation (\ref{schroed}) for various input parameters $|\mathbf{k}|$ using the potential (\ref{time-dep-pot}) computed from the same four point correlation function).
Figure \ref{phase_compare_wall_dirichlet_antiper} demonstrates that the dependence of the phase shift on the temporal boundary conditions is also negligible. This result is important when dynamical fermions are used, since anti-periodic boundary conditions are the canonical choice in that case.

\subsection{Ground state saturation}
The chosen $t_\mathrm{min}/a{=}15$ roughly corresponds to $t_\mathrm{min}{\approx} 1.7\,\mathrm{fm}$ which might also be reachable in nucleon-nucleon systems. The scattering phases obtained from L\"uscher's method show tiny deviations for different $t_\mathrm{min}$, whereas those from the potential method are barely affected since the ground state saturation is not required here. 
Indeed, the scattering phase at low energy from the potential method is stable for  even smaller $t$ as long as $t/a \ge t_{\rm inelastic}/a \approx 11$ (cf. Figure \ref{fig:t-dependence}). A variation of the scattering phase at $t {<} t_{\rm inelastic}$ is a sign of inelastic contributions.

There exists a truncation error in the derivative expansion of the potential.
The most reliable way to estimate this systematic error is to determine
higher order potential(s) explicitly \cite{Murano:2011nz,Murano:2013xxa}
and study the size of the higher order corrections.
This procedure, however, is quite demanding computationally.
We therefore employ an approximate estimate as follows:
In the region $t_{\rm inelastic}{<}t{<}t_{\rm ground}$ where inelastic
contribution can be neglected while the ground state saturation is not achieved,
a mixture of various elastic states exists in $R({\bf r},t,t_0)$ depending on $t-t_0$.
Although the original non-local potential is $t$-independent in this regions,
the truncation of the derivative expansion introduces $t$-dependent artifact.
Therefore, if higher order terms in the derivative expansion is sizeable,
the $t$ dependence of the local potentials becomes visible and may be used
as an estimate of the truncation error.
As discussed above, the $t$ dependence is almost invisible for the scattering phases at low energy,
but we nonetheless estimate and include this error in our final error-budget for the scattering length.
In general, the size of truncation error is dependent on the energy of the system.
Therefore, we estimate the truncation error in phase shifts at higher energies as well,
and find that the error is well under control, e.g., ${\sim}0.25$ \% at $E_{\rm CM}{=}150\,\mathrm{MeV}$
and ${\sim}0.13$ \% at $E_{\rm CM}{=}300\,\mathrm{MeV}$.


\subsection{Low energy parameters}

\begin{figure}
\centering
\includegraphics[scale=0.87]{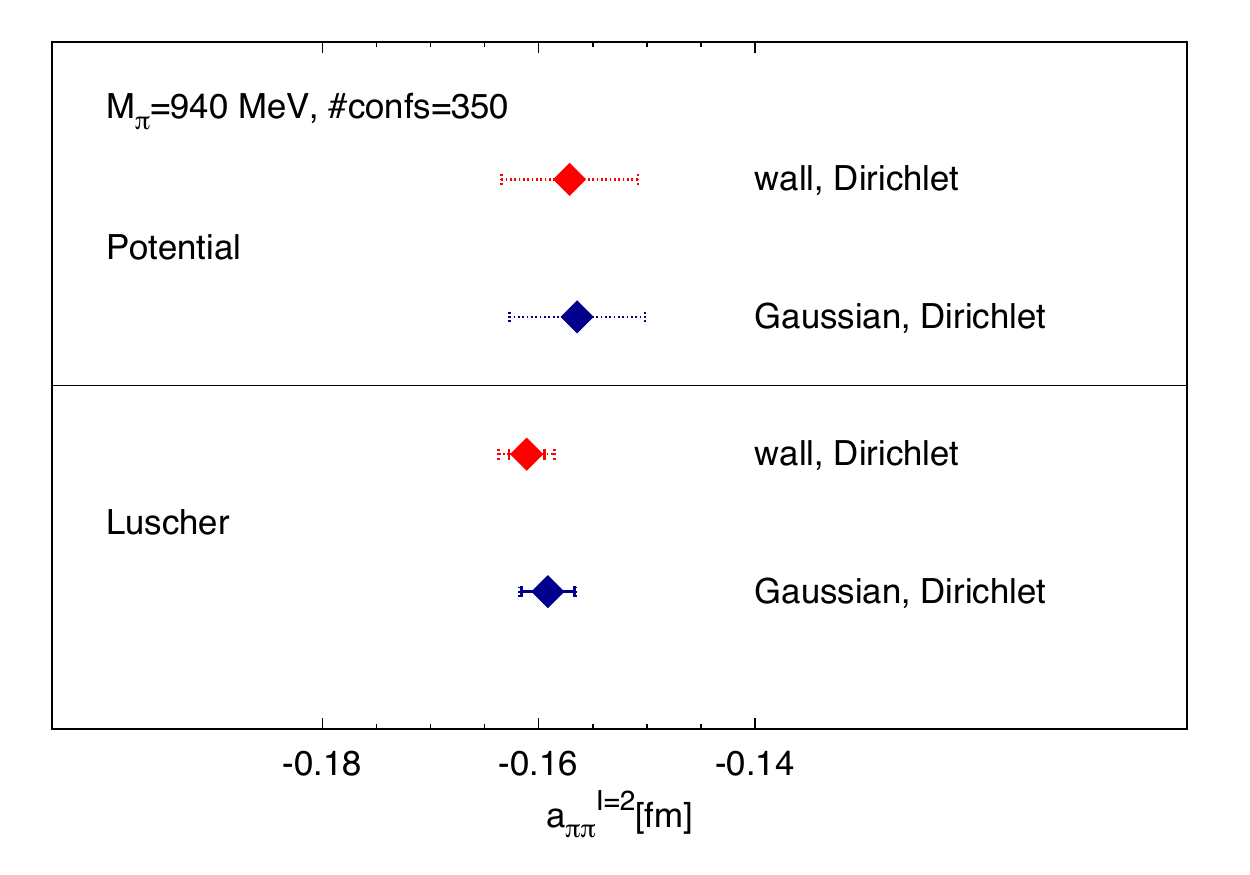}
\caption{\label{scatlengths}Results for the scattering lengths obtained using different sources and methods. The upper three results are obtained by fitting the data to the effective range expansion (\ref{ere}), the lowest one by solving the equation (\ref{deltaE0}), since we only performed the analysis for the Gaussian source on a single volume. In all cases, the statistical and overall uncertainties are displayed with solid and dotted error bars, respectively. In the upper panel the statistical error bars are smaller than the symbols. For Lüscher's method, the systematic error is underestimated (cf. section \ref{error-treatment}).}
\end{figure}

\begin{table}[!ht]
\centering
\begin{tabular}{|l|c|c|c||c|c|c|c|}
\hline
$a_{\pi\pi}^{I{=}2}[\mathrm{fm}]$ & value & stat. & sys. total & exc. states & rot.-inv. & asympt. & ERE \\\hline\hline
Potential &  -0.1568 & 0.0005 & 0.0063 & 0.0002 & 0.0062  & 0.0 & 0.0006  \\
L\"uscher & -0.1615 & 0.0020  & 0.0020 + ?   & 0.0017   & ? & -- & 0.0008  \\
\hline
\end{tabular}
\caption{\label{tab:results}Results for scattering lengths $a_{\pi\pi}^{I{=}2}$ obtained from either method
at $m_\pi{\sim}940\,\mathrm{MeV}$, including an error budget. The breakups include the effect of excited states (column 5), the violation of rotational invariance (column 6, not estimated for Lüscher's method, cf. section \ref{error-treatment}), the asymptotic behavior of the wave-function (column 7, applicable only for the potential method) and different orders in the effective range expansion (column 8). Due to correlations, the errors do not sum up to $100\%$ when added in square.}
\end{table}

The scattering length is a useful quantity in phenomenological applications, since many interesting phenomena in low-energy scattering theory can be described by this single number. Therefore, we compare the scattering lengths obtained from the two approaches by fitting the data to the effective range expansion (\ref{ere}). Since finite volume effects in the potential method are small, we extract the scattering length from our reference lattice with $L_s{=}3.7\,\mathrm{fm}$ in this case. For Lüscher's method however, we have to take into account all four different volumes. In case of the Gaussian source, we only analyzed the scattering phase shift on a single volume. Hence, we make use of the large volume expansion given in (\ref{deltaE0}) in order to compute $a_{\pi\pi}^{I{=}2}$ (we expect the higher orders in $1/L$ to be negligible). The resulting scattering lengths are displayed in Figure \ref{scatlengths} which shows
an excellent agreement. Table \ref{tab:results} displays the error budgets for the scattering lengths obtained from either method.
As seen from the Table, the overall uncertainty in the potential method is dominated by systematics associated with the violation of rotational symmetry.
As discussed in section~\ref{lattice-setup}, 
there exists an additional systematic error by the quenched approximation,
which is not included in the error budget.
In particular, the lack of unitarity in a quenched theory
could induce potentially uncontrolled errors in both methods.
The observation of the good agreement between two methods, however, indicates
that the error from this pathology is less significant compared to other errors,
presumably because of the heavy quark mass employed in this study.
To make a quantitative estimate of the quenching artifact,
an explicit simulation in full QCD is desirable,
but we leave it for future studies.


\section{Summary}
We have performed an $I{=}2$ $\pi\pi$ scattering study in quenched QCD with heavy pions of $m_\pi \sim 940$ MeV. In the determination of scattering phase shifts on the lattice, two different approaches have been employed with a particular emphasis on the examination of systematic uncertainties in each method:
L\"uscher's finite volume approach and the HAL QCD potential method.
The results of the phase shift and the scattering length have been found to agree well between the 
two methods. We have observed that the largest systematic uncertainty in the potential method stems from the violation of rotational invariance,
while such a systematic uncertainty is difficult to estimate in L\"uscher's method and thus has been neglected in this study. 
The systematic uncertainty attributed to the quenched approximation has not been evaluated for both methods. While Lüscher's method is sensitive to excited state contaminations, the time-dependent potential method 
can compensate a gross of these effects. This is especially important when multi-baryon systems are considered.
Furthermore, the potential approach allows for extracting the scattering phase at arbitrary momenta, as long as the energy of the system is below the inelastic threshold.

\acknowledgments
All computations were carried out on FZ J\"ulich and University of Wuppertal computing resources.
This work is supported in part by the Japanese Grant-in-Aid for Scientific Research (No. 24740146, 25287046), the Grant-in-Aid for Scientific Research on Innovative Areas (No. 2004: 20105001, 20105003), SPIRE (Strategic Program for Innovative Research) and the German DFG SFB TR 55 (Sonderforschungsbereich Transregio 55). Special thanks go to Zoltan Fodor for useful discussions and continuous support. We would also like to thank Kalman Szabo, Stefan Krieg, Stephan Dürr, Christian Hölbling, Szabolcs Borsanyi and Balint Toth for their support at different stages of the project.

\appendix

\section{Quark Mass dependence}
\label{sec:mass-dep}

\begin{figure}
\centering
\includegraphics[scale=1.0]{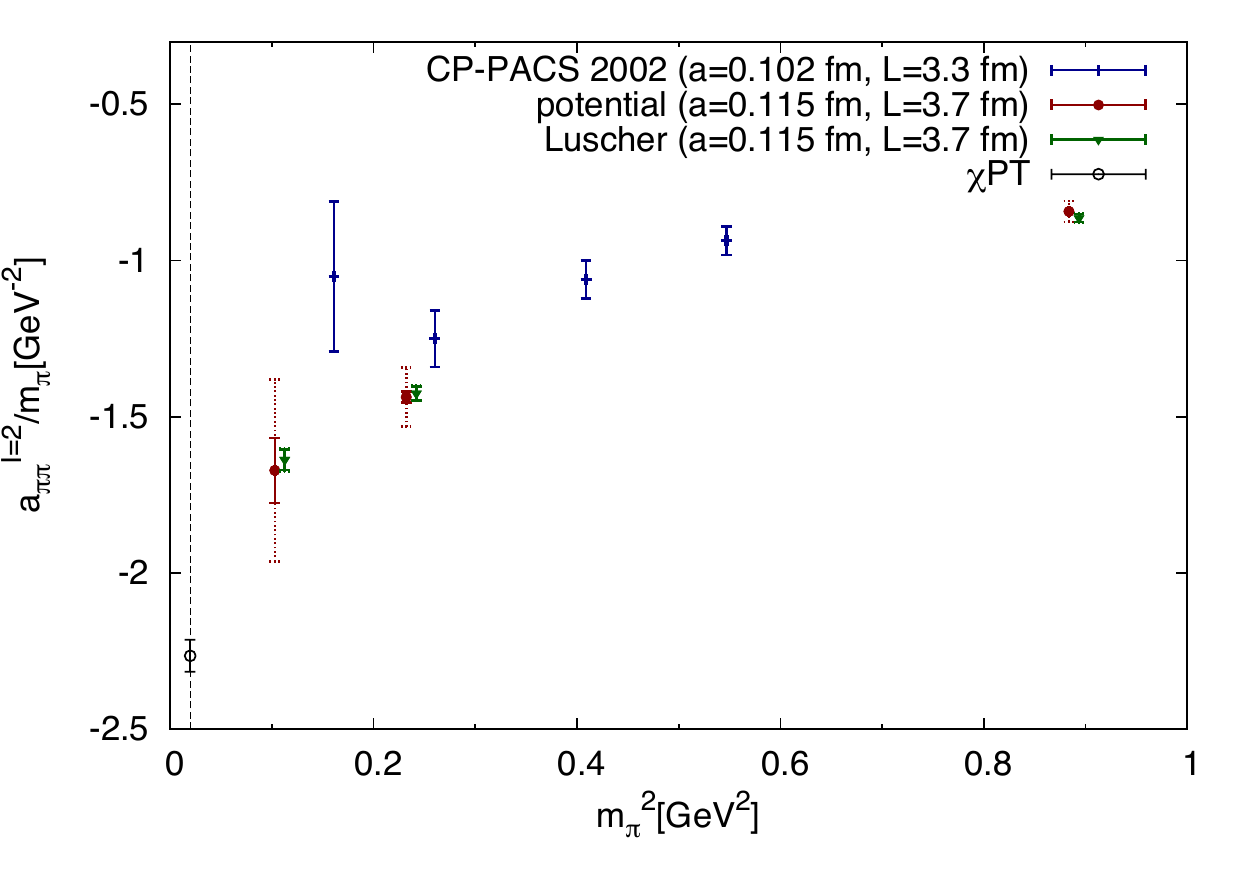}
\caption{\label{scatlengthsovermpi_vs_mpisq}Results for the scattering lengths as a function of pion mass squared. Our results from two methods (note that the results are slightly displaced in order to improve readability) are compared with results in Ref.~\protect\cite{Aoki:2002ny}.
The open black circle denotes the result obtained from continuum $\chi$PT\protect\cite{Gasser:1983kl,Bijnens:1996dz,Colangelo:2001qa}.}
\end{figure}
Systematic uncertainties in L\"uscher's approach and in the potential method could depend on pion masses. 
We make a supplemental investigation of such a dependence by lowering the pion mass 
from $m_\pi{\sim}940\,\mathrm{MeV}$ down to $m_\pi{\sim} 330\mathrm{MeV}$ and comparing the scattering lengths obtained from the two methods. 
We keep our physical volume fixed, so that finite volume effects increase when the pion mass is decreased. 
Both methods are expected to be barely affected by finite volume effects, as long as the volume remains much larger than the interaction range of the pions.
In Figure \ref{scatlengthsovermpi_vs_mpisq},
we plot the scattering lengths as a function of $m_\pi^2$.
We  see that even for pion masses as low as $m_\pi{\sim}330\,\mathrm{MeV}$, 
the results of the two methods agree very well.
As in the case of $m_\pi{\sim} 940$ MeV, the overall uncertainty in the potential method is dominated by systematics due to the violation of rotational invariance.
The results are also compared to a previous quenched calculation by CP-PACS \cite{Aoki:2002ny}, in which the lattice spacing, volume and pion masses are similar to the parameters in this work.
Figure \ref{scatlengthsovermpi_vs_mpisq} also shows the physical (continuum) value for $a_{\pi\pi}^{I{=}2}/m_\pi$ estimated from $\chi$PT \cite{Gasser:1983kl,Bijnens:1996dz,Colangelo:2001qa}. Note that the behavior $a_{\pi\pi}^{I{=}2}{\sim} 1/m_\pi$, predicted by q$\chi$PT \cite{Bernard:1995ez} is not visible, neither in our nor in the CP-PACS data. 
This might be attributed to lattice artifacts induced by the large lattice spacing (cf. \cite{Aoki:2002ny}) and the use of the unimproved plaquette action. Another possibility is that the quark masses used in this study are still too heavy to see the q$\chi$PT effect.

\FloatBarrier

\bibliographystyle{JHEP}
\bibliography{potential}

\end{document}